\documentclass[runningheads]{llncs}
\usepackage[T1]{fontenc}
\usepackage{graphicx}
\usepackage{booktabs}
\usepackage{amsmath}
\usepackage{tikz}
 
\usepackage{amssymb}
\usepackage{hyperref}
\usepackage{float}

\begin{document}
\title{STACT-Time: \textbf{S}patio-\textbf{T}emporal Cross \textbf{A}ttention for \textbf{C}ine \textbf{T}hyroid Ultrasound \textbf{Time }Series Classification}

\author{Irsyad Adam\inst{1}
\and Tengyue Zhang\inst{1,2}
\and Shrayes Raman\inst{2}
\and Zhuyu Qiu\inst{1}
\and Brandon Taraku\inst{2}
\and Hexiang Feng\inst{2}
\and Sile Wang\inst{2}
\and Ashwath Radhachandran\inst{2}
\and Shreeram Athreya\inst{3}
\and Vedrana Ivezic\inst{1}
\and Peipei Ping\inst{1}
\and Corey Arnold\inst{1,2,3,4}
\and William Speier\inst{1,2,4}
}
\authorrunning{I. Adam et al.}
%
\institute{Medical \& Imaging Informatics, Department of Radiological Sciences, David Geffen School of Medicine, University of California, Los Angeles (UCLA), CA, USA \and
Department of Bioengineering, UCLA, Los Angeles, CA, USA \and 
Department of Electrical and Computer Engineering, UCLA, Los Angeles, CA, USA
\and
Department of Radiological Sciences, UCLA, Los Angeles, CA, USA
}

\maketitle
\begin{abstract}
Thyroid cancer is among the most common cancers in the United States. Thyroid nodules are frequently detected through ultrasound (US) imaging,  and some of the nodules require further evaluation via fine-needle aspiration (FNA) biopsy. Despite its effectiveness, FNA often leads to unnecessary biopsies of benign nodules, causing patient discomfort and anxiety. To address this, the American College of Radiology Thyroid Imaging Reporting and Data System (TI-RADS) has been developed to reduce benign biopsies. However, such systems are limited by interobserver variability. Recent deep learning approaches have sought to improve risk stratification, but they often fail to utilize the rich temporal and spatial context provided by US cine clips, as cine clips contain dynamic global information and surrounding structural changes across various views. In this work, we propose the \textbf{S}patio-\textbf{T}emporal Cross \textbf{A}ttention for \textbf{C}ine \textbf{T}hyroid Ultrasound \textbf{Time }Series Classification (STACT-Time) model, a novel representation learning framework that integrates imaging features from US cine clips with features from segmentation masks automatically generated by a pretrained model. By leveraging self-attention and cross-attention mechanisms, our model captures the rich temporal and spatial context of US cine clips while enhancing feature representation through segmentation-guided learning. Our model improves malignancy prediction compared to state-of-the-art models, achieving a cross-validation precision of 0.91 ± 0.02 and F1 score of 0.89 ± 0.02. By reducing unnecessary biopsies of benign nodules while maintaining high sensitivity for malignancy detection, our model has the potential to enhance clinical decision-making and improve patient outcomes.

\end{abstract}
\section{Introduction}
Thyroid cancer is one of the most frequently diagnosed cancers in the United States~\cite{siegel2024cancer}. Thyroid nodules are small lumps that can form in the thyroid gland. While most of them are benign, some malignant thyroid nodules can lead to cancer. Clinically, ultrasound (US) imaging is used for the initial assessment of thyroid nodules due to its ability to visualize the gland and nodules, allowing clinicians to assess the echogenicity, composition, shape, margins, and other sonographic features~\cite{aium2013guideline}. However, the findings of US imaging are often nonspecific; a nodule classified as suspicious may undergo fine-needle aspiration (FNA) biopsy to determine whether it is benign or may require surgery~\cite{HOANG2015143}. Since most nodules are benign, many patients ultimately receive a benign diagnosis after the FNA biopsy. Despite FNA's minimal invasiveness, it can still cause discomfort and anxiety to patients particularly when multiple biopsies are required. To reduce unnecessary biopsies while effectively identifying clinically significant cancers, the American College of Radiology (ACR) published the thyroid ultrasound lexicon white paper and formed the basis for the ACR Thyroid Imaging Reporting and Data System (TI-RADS) for risk stratification of thyroid nodules~\cite{tessler2017acr}, which has been shown to reduce the number of unnecessary biopsies~\cite{grani2019reducing}. Nonetheless, a notable challenge in this approach is the interobserver variability in interpreting sonographic features, which can lead to inconsistencies in clinical management recommendations. 

\begin{figure}[H]
\center
\includegraphics[width=0.7\textwidth]{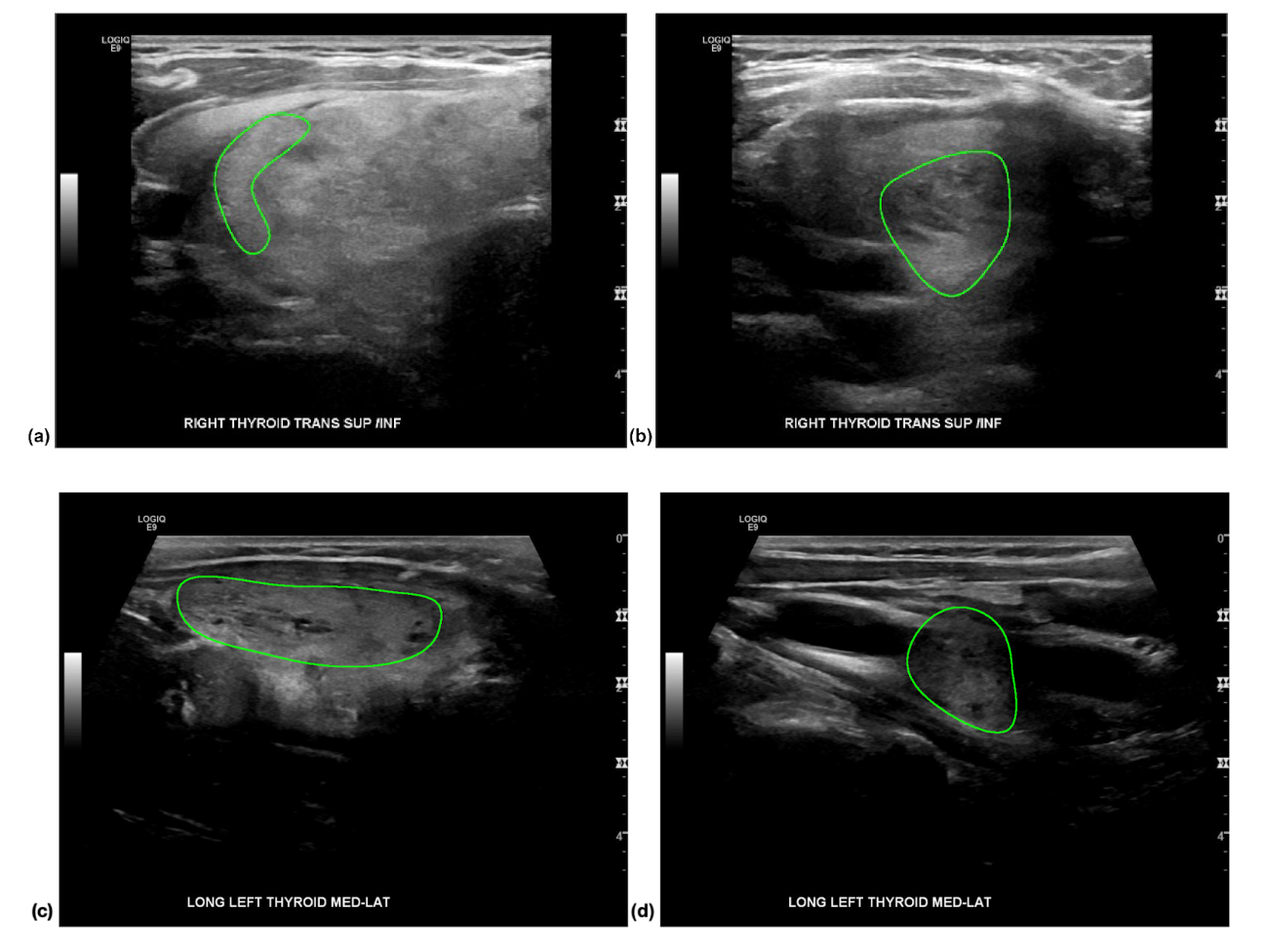}
\caption{An example showing annotated ROIs of identified nodules during a sweep by the scanning scope. Example cine clips of the first (a, c) and the last (b, d) frames of two representative subjects with the nodule contour (green line) overlaid.
} \label{fig:case_study}
\end{figure}

To address the above problems, existing research has aimed to reduce false positives in US imaging. Many studies have focused on improving risk stratification from US imaging using deep learning approaches~\cite{buda2019management,zhuang2023patient,radhachandran2024thygraph,zhou2020differential,stanford}. Moreover, recent works have explored incorporating nodule boundary information within a single US image frame into the malignancy classification task~\cite{yao2023human,wang2024interpretable}. However, these approaches are limited to static frames and do not utilize the rich temporal and spatial contexts provided by cine (video) clips. Clinically, cine clips are collected using real-time, continuous scanning of a target nodule, capturing tissue structure, global changes, and multiple views, overall providing rich temporal and spatial contexts. As illustrated in Fig.~\ref{fig:case_study}, the beginning and end frames of a cine clip often provide complementary information about the nodule's overall tissue deformation and contour changes. While a standard vision model attempts to incorporate these signals into an embedding, segmentation masks are directly able to capture these tissue shapes by visualizing the boundary of the region of interest. Furthermore, while there are protocols to guide the views of the cine recording, these recordings are often operator-dependent, and trajectories may change based on discretion to optimize nodule visibility. Additionally, thyroid ultrasound clips lack stable anatomical landmarks for precise movement, and the surrounding tissue of the nodule may shift with patient activity (i.e. breathing and swallowing). All of these factors make it difficult to establish consistent spatial correspondences across cine frames, as seen in Fig. \ref{fig:case_study}. However, each frame from the cine clip gives enhanced contextual understanding by displaying various views of the shape and size of the same nodule from the perspective of different angles. Therefore, in order to effectively incorporate the spatio-temporal information provided by nodule segmentation into malignancy classification, we incorporate cross-attention to systematically enrich the imaging features from each frame with structural information from the nodule segmentation across all frames. 
In this work, we propose a \textbf{S}patio-\textbf{T}emporal Cross \textbf{A}ttention for Segmentation Augmented \textbf{C}ine Ultrasound \textbf{T}hyroid \textbf{Time} Series Classification (STACT-Time) model, a novel framework for thyroid nodule malignancy classification that effectively integrates imaging features and segmentation features in time-series cine clip data. In STACT-Time, the self-attention mechanism was used to capture spatial and temporal relationships within the original image modality. The cross-attention mechanism was used to integrate the segmentation views into the imaging modality for segmentation-augmented representation. In other domains, cross-attention mechanisms have demonstrated effectiveness in tasks such as multimodal learning for image-text matching~\cite{wei2020multi}, multimodal medical image registration~\cite{song2021cross}, and integrating heterogeneous data sources for medical decision making~\cite{kim2024llm}. Leveraging these mechanisms, our framework effectively learns a representation of cine clips by capturing rich temporal and spatial information in US images while enriching it by the contextual information provided by nodule segmentation. This segmentation-aware approach significantly enhances model performance, achieving improved accuracy and precision compared to models that did not utilize segmentation information.

\section{Method}

\subsection{Dataset}

\begin{table}[t]
\caption{Summary characteristics of the study cohort ($N$=192 nodules) with comparison between malignant and benign nodules. All statistics were calculated on the nodule level. P-values were calculated using the Mann-Whitney U Test.}
\label{tab:dataset}
\centering
\renewcommand{\arraystretch}{0.9}
\setlength{\tabcolsep}{6pt}
\begin{tabular}{p{2.8cm}p{1.9cm}p{1.9cm}p{2.1cm}p{1.3cm}}
\toprule
\textbf{Variables} & \textbf{All} & \textbf{Benign} & \textbf{Malignant} & \textbf{P-value} \\
\midrule
Number of nodules & 192 (100\%) & 175 (91.15\%) & 17 (8.85\%)  & - \\
Age & 56.04 ± 56.04 & 56.79 ± 56.79 & 48.29 ± 48.29 & 0.02 \\
Female & 159 (82.81\%) & 144 (82.29\%) & 15 (88.24\%) & - \\
\midrule
\textbf{Location} & & & & \\
\multicolumn{1}{r}{Left lobe} & 72 (37.50\%) & 65 (37.14\%) & 7 (41.18\%) & - \\
\multicolumn{1}{r}{Right lobe} & 108 (56.25\%) & 100 (57.14\%) & 8 (47.06\%) & - \\
\multicolumn{1}{r}{Isthmus} & 12 (6.25\%) & 10 (5.71\%) & 2 (11.76\%) & - \\
\midrule
\textbf{TI-RADS} & & & & \\
\multicolumn{1}{r}{TR1} & 1 (0.52\%) & 1 (0.57\%) & 0 (0\%) & - \\
\multicolumn{1}{r}{TR2} & 10 (5.21\%) & 10 (5.71\%) & 0 (0\%) & - \\
\multicolumn{1}{r}{TR3} & 52 (27.08\%) & 52 (29.71\%) & 0 (0\%) & - \\
\multicolumn{1}{r}{TR4} & 83 (43.23\%) & 78 (44.57\%) & 5 (29.41\%) & - \\
\multicolumn{1}{r}{TR5} & 46 (23.96\%) & 34 (19.43\%) & 12 (70.59\%) & - \\
\midrule
Number of frames  & 90.69 ± 56.05 & 88.25 ± 55.61 & 115.76 ± 54.39 & 0.03 \\
\bottomrule
\end{tabular}
\end{table}

We used the Stanford CINE dataset in this study~\cite{stanford}. This dataset contains $N=192$ nodules from Stanford University School of Medicine. For each nodule, US cine-clip images, radiologist-annotated nodule ROI(s), patient demographics, lesion size and location, TI-RADS descriptors, and biopsy-confirmed diagnoses were provided. On average, there are 91 frames per nodule. All US imaging was reviewed by board-certified radiologists for quality. Thyroid nodules were accurately segmented using the semi-automated Electronic Physician Annotation Device (ePADlite) software by experienced radiologists. See Table \ref{tab:dataset} for a summary of the dataset characteristics.

\subsection{Model Architecture}

\begin{figure}[t]
\includegraphics[width=\textwidth]{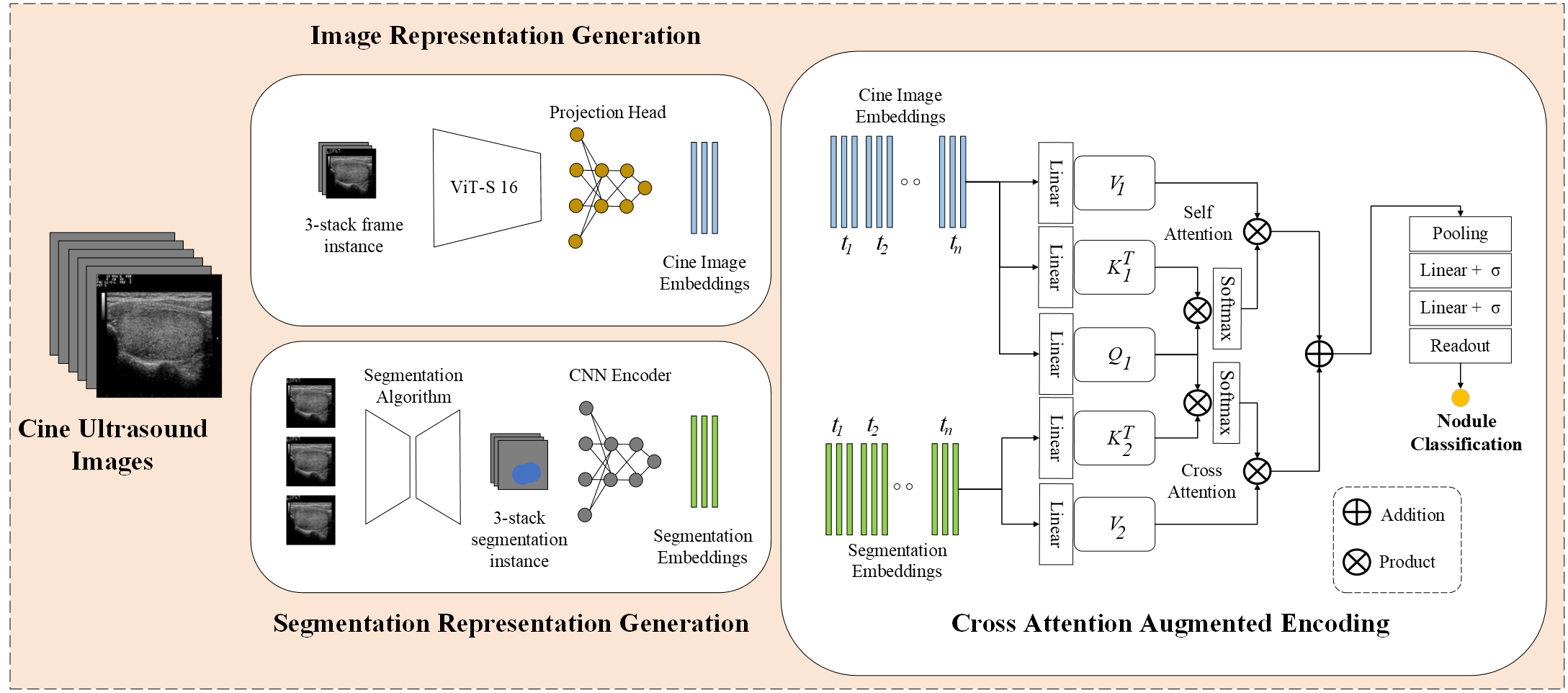}
\caption{Overview of our proposed segmentation-augmented time-series representation learning framework.} \label{fig:overview}
\end{figure}

Fig. \ref{fig:overview} shows the overview of the proposed segmentation-augmented time-series representation learning framework for thyroid nodule malignancy prediction. Cine image embeddings were refined using self-attention to capture internal dependencies and were enhanced by segmentation embeddings using cross-attention to integrate complementary information from the segmentation masks. Representations obtained from each attention module were concatenated, and the combined features were aggregated through pooling and passed to a classification head to predict nodule malignancy.

\subsection{Feature Encoders}

\noindent\textbf{Image Encoder.} The ViT model pre-trained on ImageNet~\cite{vit} was used as the image encoder. Features were extracted from a bounding box around each nodule, with a 5-pixel buffer around the nodule region of interest (ROI). We finetuned the last two fully connected layers of the ViT model using US cine clips and nodule malignancy labels. Since ViT requires 3-channel RGB images, following the approach in~\cite{stanford}, we stacked three adjacent frames from a US cine clip to form a single 3-channel image input (3-stack frame instance). Each cine clip was transformed into multiple 3-stacks using a non-overlapping window approach. The label for the nodule was assigned to each 3-stack frame instance capturing this nodule. After fine-tuning, a 256-dimensional feature vector was extracted from the last hidden layer. This results in an image embedding $I=[t_1, t_2, ..., t_{n/3}] \in \mathbb{R}^{(n/3)\times 256}$ for each cine clip with $n$ frames. 

\noindent\textbf{Segmentation Encoder.} To generate segmentation embeddings, we used the MSUnet architecture~\cite{su2021msu} trained on thyroid images obtained from **** [*]. The pre-trained MSUnet was then fine-tuned on our US cine dataset with nodule ROI annotations. The fine-tuned model produced binary segmentation masks for each frame, allowing the model to perform inference with only cine clips without requiring radiologist-annotated nodule ROIs. Similar to the image encoder, segmentation masks of adjacent frames were stacked into a 3-stack segmentation instance. The 3-stack segmentation instances were then encoded using a 2-layer convolutional neural network (CNN) trained on nodule malignancy labels to predict histopathological diagnosis. 256 features from the last hidden layer were extracted as the stack-level segmentation embeddings. This results in a segmentation embedding $S=[t'_1, t'_2, ..., t'_{n/3}] \in \mathbb{R}^{(n/3) \times 256}$ for each cine clip with $n$ frames. 

\subsection{Cross Attention Augmented Encoding}
\noindent\textbf{Self-Attention Mechanism.} For the thyroid images, a self-attention mechanism is applied to capture relationships within the cine clip. A sinusoidal positional embedding was initially applied to both the image and the segmentation embeddings to retain temporal information. Then, the image embeddings are first projected into query ($Q_1$), key ($K_1$), and value ($V_1$) spaces by learnable linear projections. For each embedding, attention scores are computed with every other embedding in the sequence by calculating a dot product between $Q$ and $K$ and normalizing through a softmax function. The final attention scores have the same shape as the input, with each embedding enriched by contextual information from the entire sequence.

\noindent\textbf{Cross-Attention Mechanism.} To incorporate information across cine image and segmentation embeddings, a cross-attention module was used to guide the cine image embeddings with the segmentation embeddings. In the cross-attention mechanism, embeddings from one modality are chosen as the query, and the other modality provides key and value. Segmentation embeddings were chosen as the key ($K_2$) and value ($V_2$) to provide structural and spatial guidance on the image embeddings. US image embeddings were chosen as the query ($Q_2$) because they capture more complex, dynamic features of the cine clips. 

\subsection{Implementation Details}
The method was implemented in Pytorch, using an NVIDIA A40 GPU with 48 GB of memory. The ViT backbone was fine-tuned for 100 epochs. Stochastic gradient descent (SDG) was used as the optimizer with an initial learning rate of  0.0001 and 0.9 momentum. A CosineAnnealingLR scheduler was used to handle lr adjustments. To tackle the class imbalance problem, a focal loss was used with $\alpha = 0.9$, $\gamma = 2.4$. The MSUnet was fine-tuned from the respective checkpoint for 50 epochs, and the training loop was replicated from [*]. For the cross-attention module, the decoder was trained for 100 epochs, using a CosineAnnealingLR and AdamW optimizer with initial learning rate of  $0.0001$. Focal loss was used to accomodate for the class imbalance.


\section{Results}

\noindent\textbf{Comparison with Existing Methods.} We compared our model with three state-of-the-art cine clip-based nodule malignancy models: CineCNNAvePool~\cite{stanford}, Cine-CNNTrans~\cite{stanford}, and the TI-RADS descriptors. CineCNNAvePool aggregated stack-level features into nodule level using average pooling. Cine-CNNTrans further improved the CineCNNAvePool model by using a transformer encoder to perform feature aggregation, achieving an average AUC of 0.88 ± 0.10. The TI-RADS model is a random forest model using five TI-RADS descriptors (composition, echogenicity, shape, margin, and punctate echogenic foci) and one TI-RADS score as input to predict histopathological diagnosis. In comparison, our model achieved a higher AUC of 0.91 ± 0.01, outperforming previous models while also demonstrating superior precision (0.91 ± 0.02) and F1 score (0.89 ± 0.02), indicating its ability to accurately identify malignant nodules while maintaining a balance between sensitivity and specificity. 

\begin{table}[h!]
\caption{Comparison with existing methods for nodule malignancy classification using cine clips. Performance is reported as mean ± standard deviation over 5-fold cross-validation. Best values are bolded. }
\label{tab:compare_existing}
\centering
\setlength{\tabcolsep}{6pt}
\resizebox{\textwidth}{!}{%
\begin{tabular}{p{3cm}p{1.8cm}p{1.8cm}p{1.8cm}p{1.8cm}p{1.8cm}}
\toprule
& \textbf{Nodule ACC} & \textbf{Nodule Precision} & \textbf{Nodule Recall} & \textbf{Nodule F1} & \textbf{Nodule AUROC} \\
\midrule
Cine-CNNAvePool & - & - & - & - & 0.82 ± 0.07 \\
Cine-CNNTrans & - & - & - & - & 0.88 ± 0.10 \\
TI-RADS & \textbf{0.90 ± 0.02} & 0.86 ± 0.06 & \textbf{0.90 ± 0.02} & 0.88 ± 0.04 & 0.84 ± 0.08 \\
Segmentation + \\Transformer & 0.80 ± 0.02 & 0.63± 0.03 & 0.80 ± 0.02 & 0.71 ± 0.01 & 0.82 ± 0.03 \\ 
ViT + Transformer &  0.87 ± 0.01 & 0.83 ± 0.02 & 0.75± 0.02 & 0.75± 0.3 & 0.87 ± 0.02 \\ 
STACT-Time & 0.89 ± 0.03 & \textbf{0.91 ± 0.02} & 0.88 ± 0.01 & \textbf{0.89 ± 0.02} & \textbf{0.91 ± 0.01} \\
\bottomrule
\end{tabular}}
\end{table}

\noindent\textbf{Ablation Studies.} Ablation studies were conducted to evaluate the contribution of each module. The performance of various backbone architectures on frame- and nodule-level classification tasks is presented in Table~\ref{tab:backbone}. ShuffleNet v2~\cite{shufflenetv2} and MobileNet v2~\cite{mobilenetv2} demonstrated comparable performance, with frame-level ACC of 0.88 and nodule-level AUROC around 0.78-0.79. The ViT backbone achieved higher frame ACC (0.92), frame AUROC (0.76), and nodule AUROC (0.83). The segmentation backbone, though showing promising results, had comparatively lower performance (nodule ACC=0.90, nodule AUROC=0.80) due to the lack of textural information in the input binary masks. The Joint Fusion backbone, where the image and segmentation embeddings were concatenated, achieved higher frame-level ACC (0.938±0.02), AUROC (0.830±0.03), and nodule-level AUROC (0.832±0.03) than all other backbones. This validates the effectiveness of using segmentation embeddings to enrich image embeddings for nodule malignancy classification. 

We further compared the nodule-level classification performance of the proposed model with a model using the ViT backbone with self-attention and one using the segmentation backbone with self-attention, as shown in Table~\ref{tab:nodule_classification}. By enriching the image embeddings with contextual information from the segmentation embeddings, the proposed model increased nodule-level ACC by 0.02, precision by 0.08, and F1 score by 0.19 compared to the ViT + self-attention approach. Additionally, it outperformed the segmentation + self-attention model with improvements of 0.09 in nodule-level ACC, 0.28 in precision, and 0.18 in F1 score. 

\begin{table}[t]
\caption{Accuracy Metrics on Various Backbones. }
\label{tab:backbone}
\centering
\setlength{\tabcolsep}{6pt}
\resizebox{\textwidth}{!}{%
\begin{tabular}{p{3.4cm}p{1.8cm}p{1.8cm}p{1.8cm}p{1.8cm}}
\toprule
\textbf{} & \textbf{Frame ACC} & \textbf{Nodule ACC} & \textbf{Frame AUROC} & \textbf{Nodule AUROC} \\
\midrule
ShuffleNet v2 Backbone & 0.88 ± 0.01 & \textbf{0.88 ± 0.01} & 0.69 ± 0.02 & 0.79 ± 0.03\\
MobileNet v2 Backbone & 0.88 ± 0.01 & \textbf{0.88 ± 0.01} & 0.73 ± 0.02 & 0.78 ± 0.01 \\
ViT Backbone & 0.92 ± 0.01 & 0.81 ± 0.01 & 0.76 ± 0.03 & 0.83 ± 0.05 \\
Segmentation Backbone & 0.90 ± 0.02& 0.79 ± 0.02& 0.74 ± 0.03& 0.80± 0.02\\
Joint Fusion Backbone & \textbf{0.94 ± 0.02} & 0.85 ± 0.01 & \textbf{0.83 ± 0.03} & \textbf{0.83 ± 0.03} \\
\bottomrule
\end{tabular}}
\end{table}

\begin{table}[t]
\caption{Comparison of Attention Mechanisms for Nodule-Wise Time Series Classification. ``Self'': self attention. ``Cross'': cross attention. ``Seg'': segmentation.}\label{tab:nodule_classification}
\centering
\setlength{\tabcolsep}{6pt}
\resizebox{\textwidth}{!}{%
\begin{tabular}{p{2.9cm}p{1.8cm}p{1.8cm}p{1.8cm}p{1.8cm}p{1.8cm}}
\toprule
\textbf{} & \textbf{ACC} & \textbf{Precision} & \textbf{Recall} & \textbf{F1} & \textbf{AUROC} \\
\midrule
ViT + Self & 0.873 ± 0.01
 & 0.83 ± 0.02 & 0.76 ± 0.02 & 0.76 ± 0.3 & 0.87 ± 0.02\\
Seg+ Self& 0.80 ± 0.02 & 0.636 ± 0.03 & 0.806 ± 0.02 & 0.711 ± 0.01 & 0.823 ± 0.03 \\
ViT+Seg+Cross (STACT-Time) & \textbf{0.89 ± 0.03}&\textbf{ 0.91± 0.02}& \textbf{0.88 ± 0.01} & \textbf{0.89 ± 0.02}& \textbf{0.91 ± 0.01}\\
\bottomrule
\end{tabular}}
\end{table}

\section{Discussion}

Compared with prior models, our approach outperformed state-of-the-art methods, including CineCNNAvePool, Cine-CNNTrans, and TI-RADS descriptors, in thyroid nodule malignancy classification. By integrating segmentation embeddings with the cross-attention mechanism, our model achieved higher AUC, precision, and F1 scores, highlighting its ability to balance sensitivity and specificity effectively. These results underscore the importance of combining the spatio-temporal contextual information from segmentation in cine clips to enhance malignancy prediction. 

The ablation study demonstrated the individual contributions of different modules to the overall performance. While the ViT backbone captures imaging features and shows relatively high accuracy in nodule-level classification, it lacks the ability to capture contextual information for precise prediction. While the segmentation backbone was promising, it had low precision and F1 scores, indicating less predictive ability due to the lack of imaging details from binary segmentation masks alone. In contrast, the proposed STACT-Time model successfully combined these complementary features, achieving the highest metrics across frame- and nodule-level classification tasks. These findings validate the effectiveness of our approach in addressing the limitations of existing methods that did not take nodule segmentation across the entire cine clips into account. 

In summary, this study proposes a Spatio-Temporal Cross Attention for Segmentation Augmented Cine Ultrasound Thyroid Time Series Classification framework (STACT-Time). By leveraging self-attention and cross-attention mechanisms, our model effectively integrates the imaging and contextual information provided by the cine clips and the location and structural information from the nodule segmentation masks. Experimental results demonstrate that our model outperforms conventional TI-RADS risk stratification systems and existing machine learning approaches, offering a more accurate nodule risk stratification model that holds the potential to reduce false positives in US imaging and avoid unnecessary FNA biopsies. Future work will validate our model on larger, multi-institutional external datasets to validate its generalizability. 

\begin{credits}
\subsubsection{\ackname} 
\subsubsection{\discintname}
\end{credits}

\bibliographystyle{splncs04}
\bibliography{references}

\begin{thebibliography}{10}
\providecommand{\url}[1]{\texttt{#1}}
\providecommand{\urlprefix}{URL }
\providecommand{\doi}[1]{https://doi.org/#1}

\bibitem{aium2013guideline}
{American Institute of Ultrasound in Medicine}, {American College of Radiology}, {Society for Pediatric Radiology}, {Society of Radiologists in Ultrasound}: {AIUM} practice guideline for the performance of a thyroid and parathyroid ultrasound examination. Journal of Ultrasound in Medicine  \textbf{32}(7),  1319--1329 (July 2013). \doi{10.7863/ultra.32.7.1319}

\bibitem{buda2019management}
Buda, M., Wildman-Tobriner, B., Hoang, J.K., Thayer, D., Tessler, F.N., Middleton, W.D., Mazurowski, M.A.: Management of thyroid nodules seen on us images: deep learning may match performance of radiologists. Radiology  \textbf{292}(3),  695--701 (2019)

\bibitem{vit}
Dosovitskiy, A.: An image is worth 16x16 words: Transformers for image recognition at scale. arXiv preprint arXiv:2010.11929  (2020)

\bibitem{grani2019reducing}
Grani, G., Lamartina, L., Ascoli, V., Bosco, D., Biffoni, M., Giacomelli, L., Maranghi, M., Falcone, R., Ramundo, V., Cantisani, V., et~al.: Reducing the number of unnecessary thyroid biopsies while improving diagnostic accuracy: toward the “right” tirads. The Journal of Clinical Endocrinology \& Metabolism  \textbf{104}(1),  95--102 (2019)

\bibitem{HOANG2015143}
Hoang, J.K., Langer, J.E., Middleton, W.D., Wu, C.C., Hammers, L.W., Cronan, J.J., Tessler, F.N., Grant, E.G., Berland, L.L.: Managing incidental thyroid nodules detected on imaging: White paper of the acr incidental thyroid findings committee. Journal of the American College of Radiology  \textbf{12}(2),  143--150 (2015). \doi{https://doi.org/10.1016/j.jacr.2014.09.038}, \url{https://www.sciencedirect.com/science/article/pii/S1546144014006279}

\bibitem{kim2024llm}
Kim, K., Lee, Y., Park, D., Eo, T., Youn, D., Lee, H., Hwang, D.: Llm-guided multi-modal multiple instance learning for 5-year overall survival prediction of lung cancer. In: International Conference on Medical Image Computing and Computer-Assisted Intervention. pp. 239--249. Springer (2024)

\bibitem{shufflenetv2}
Ma, N., Zhang, X., Zheng, H.T., Sun, J.: Shufflenet v2: Practical guidelines for efficient cnn architecture design. In: Proceedings of the European conference on computer vision (ECCV). pp. 116--131 (2018)

\bibitem{radhachandran2024thygraph}
Radhachandran, A., Vittalam, A., Ivezic, V., Sant, V., Athreya, S., Moleta, C., Patel, M., Masamed, R., Arnold, C., Speier, W.: Thygraph: A graph-based approach for thyroid nodule diagnosis from ultrasound studies. In: International Conference on Medical Image Computing and Computer-Assisted Intervention. pp. 753--763. Springer (2024)

\bibitem{mobilenetv2}
Sandler, M., Howard, A., Zhu, M., Zhmoginov, A., Chen, L.C.: Mobilenetv2: Inverted residuals and linear bottlenecks. In: Proceedings of the IEEE conference on computer vision and pattern recognition. pp. 4510--4520 (2018)

\bibitem{siegel2024cancer}
Siegel, R.L., Giaquinto, A.N., Jemal, A.: Cancer statistics, 2024. CA: a cancer journal for clinicians  \textbf{74}(1) (2024)

\bibitem{song2021cross}
Song, X., Guo, H., Xu, X., Chao, H., Xu, S., Turkbey, B., Wood, B.J., Wang, G., Yan, P.: Cross-modal attention for mri and ultrasound volume registration. In: Medical Image Computing and Computer Assisted Intervention--MICCAI 2021: 24th International Conference, Strasbourg, France, September 27--October 1, 2021, Proceedings, Part IV 24. pp. 66--75. Springer (2021)

\bibitem{su2021msu}
Su, R., Zhang, D., Liu, J., Cheng, C.: Msu-net: Multi-scale u-net for 2d medical image segmentation. Frontiers in Genetics  \textbf{12},  639930 (2021)

\bibitem{tessler2017acr}
Tessler, F.N., Middleton, W.D., Grant, E.G., Hoang, J.K., Berland, L.L., Teefey, S.A., Cronan, J.J., Beland, M.D., Desser, T.S., Frates, M.C., et~al.: Acr thyroid imaging, reporting and data system (ti-rads): white paper of the acr ti-rads committee. Journal of the American college of radiology  \textbf{14}(5),  587--595 (2017)

\bibitem{wang2024interpretable}
Wang, M., Chen, C., Xu, Z., Xu, L., Zhan, W., Xiao, J., Hou, Y., Huang, B., Huang, L., Li, S.: An interpretable two-branch bi-coordinate network based on multi-grained domain knowledge for classification of thyroid nodules in ultrasound images. Medical Image Analysis  \textbf{97},  103255 (2024)

\bibitem{wei2020multi}
Wei, X., Zhang, T., Li, Y., Zhang, Y., Wu, F.: Multi-modality cross attention network for image and sentence matching. In: Proceedings of the IEEE/CVF conference on computer vision and pattern recognition. pp. 10941--10950 (2020)

\bibitem{stanford}
Yamashita, R., Kapoor, T., Alam, M.N., Galimzianova, A., Syed, S.A., Ugur~Akdogan, M., Alkim, E., Wentland, A.L., Madhuripan, N., Goff, D., et~al.: Toward reduction in false-positive thyroid nodule biopsies with a deep learning--based risk stratification system using us cine-clip images. Radiology: Artificial Intelligence  \textbf{4}(3),  e210174 (2022)

\bibitem{yao2023human}
Yao, S., Shen, P., Dai, T., Dai, F., Wang, Y., Zhang, W., Lu, H.: Human understandable thyroid ultrasound imaging ai report system—a bridge between ai and clinicians. Iscience  \textbf{26}(4) (2023)

\bibitem{zhou2020differential}
Zhou, H., Jin, Y., Dai, L., Zhang, M., Qiu, Y., Tian, J., Zheng, J., et~al.: Differential diagnosis of benign and malignant thyroid nodules using deep learning radiomics of thyroid ultrasound images. European Journal of Radiology  \textbf{127},  108992 (2020)

\bibitem{zhuang2023patient}
Zhuang, L., Ivezic, V., Feng, J., Shen, C., Radhachandran, A., Sant, V., Patel, M., Masamed, R., Arnold, C., Speier, W.: Patient-level thyroid cancer classification using attention multiple instance learning on fused multi-scale ultrasound image features. In: AMIA Annual Symposium Proceedings. vol.~2023, p.~1344. American Medical Informatics Association (2023)

\end{thebibliography}

\end{document}